\documentclass[twocolumn]{aa}
\usepackage{graphicx,amsmath}

\usepackage{txfonts}
\newcommand{\hess}{H.E.S.S.}
\newcommand{\xte}{RXTE}
\newcommand{\pks}{PKS~2155$-$304}
\newcommand {\gamgam}{$\gamma-\gamma$ }

\begin{document}

   \title{Multi-wavelength observations of PKS 2155-304 with H.E.S.S. }

\author{F. Aharonian\inst{1}
 \and A.G.~Akhperjanian \inst{2}
 \and A.R.~Bazer-Bachi \inst{3}
 \and M.~Beilicke \inst{4}
 \and W.~Benbow \inst{1}
 \and D.~Berge \inst{1}
 \and K.~Bernl\"ohr \inst{1,5}
 \and C.~Boisson \inst{6}
 \and O.~Bolz \inst{1}
 \and V.~Borrel \inst{3}
 \and I.~Braun \inst{1}
 \and F.~Breitling \inst{5}
 \and A.M.~Brown \inst{7}
 \and P.M.~Chadwick \inst{7}
 \and L.-M.~Chounet \inst{8}
 \and R.~Cornils \inst{4}
 \and L.~Costamante \inst{1,20}
 \and B.~Degrange \inst{8}
 \and H.J.~Dickinson \inst{7}
 \and A.~Djannati-Ata\"i \inst{9}
 \and L.O'C.~Drury \inst{10}
 \and G.~Dubus \inst{8}
 \and D.~Emmanoulopoulos \inst{11}
 \and P.~Espigat \inst{9}
 \and F.~Feinstein \inst{12}
 \and G.~Fontaine \inst{8}
 \and Y.~Fuchs \inst{13}
 \and S.~Funk \inst{1}
 \and Y.A.~Gallant \inst{12}
 \and B.~Giebels \inst{8}
 \and S.~Gillessen \inst{1}
 \and J.F.~Glicenstein \inst{14}
 \and P.~Goret \inst{14}
 \and C.~Hadjichristidis \inst{7}
 \and M.~Hauser \inst{11}
 \and G.~Heinzelmann \inst{4}
 \and G.~Henri \inst{13}
 \and G.~Hermann \inst{1}
 \and J.A.~Hinton \inst{1}
 \and W.~Hofmann \inst{1}
 \and M.~Holleran \inst{15}
 \and D.~Horns \inst{1}
 \and A.~Jacholkowska \inst{12}
 \and O.C.~de~Jager \inst{15}
 \and B.~Kh\'elifi \inst{1}
 \and Nu.~Komin \inst{5}
 \and A.~Konopelko \inst{1,5}
 \and I.J.~Latham \inst{7}
 \and R.~Le Gallou \inst{7}
 \and A.~Lemi\`ere \inst{9}
 \and M.~Lemoine-Goumard \inst{8}
 \and N.~Leroy \inst{8}
 \and T.~Lohse \inst{5}
 \and J.M.~Martin \inst{6}
 \and O.~Martineau-Huynh \inst{16}
 \and A.~Marcowith \inst{3}
 \and C.~Masterson \inst{1,20}
 \and T.J.L.~McComb \inst{7}
 \and M.~de~Naurois \inst{16}
 \and S.J.~Nolan \inst{7}
 \and A.~Noutsos \inst{7}
 \and K.J.~Orford \inst{7}
 \and J.L.~Osborne \inst{7}
 \and M.~Ouchrif \inst{16,20}
 \and M.~Panter \inst{1}
 \and G.~Pelletier \inst{13}
 \and S.~Pita \inst{9}
 \and G.~P\"uhlhofer \inst{1,11}
 \and M.~Punch \inst{9}
 \and B.C.~Raubenheimer \inst{15}
 \and M.~Raue \inst{4}
 \and J.~Raux \inst{16}
 \and S.M.~Rayner \inst{7}
 \and A.~Reimer \inst{17}
 \and O.~Reimer \inst{17}
 \and J.~Ripken \inst{4}
 \and L.~Rob \inst{18}
 \and L.~Rolland \inst{16}
 \and G.~Rowell \inst{1}
 \and V.~Sahakian \inst{2}
 \and L.~Saug\'e \inst{13}
 \and S.~Schlenker \inst{5}
 \and R.~Schlickeiser \inst{17}
 \and C.~Schuster \inst{17}
 \and U.~Schwanke \inst{5}
 \and M.~Siewert \inst{17}
 \and H.~Sol \inst{6}
 \and D.~Spangler \inst{7}
 \and R.~Steenkamp \inst{19}
 \and C.~Stegmann \inst{5}
 \and J.-P.~Tavernet \inst{16}
 \and R.~Terrier \inst{9}
 \and C.G.~Th\'eoret \inst{9}
 \and M.~Tluczykont \inst{8,20}
 \and G.~Vasileiadis \inst{12}
 \and C.~Venter \inst{15}
 \and P.~Vincent \inst{16}
 \and H.J.~V\"olk \inst{1}
 \and S.J.~Wagner \inst{11}}

\institute{
Max-Planck-Institut f\"ur Kernphysik, Heidelberg, Germany
\and
 Yerevan Physics Institute, Armenia
\and
Centre d'Etude Spatiale des Rayonnements, CNRS/UPS, Toulouse, France
\and
Universit\"at Hamburg, Institut f\"ur Experimentalphysik, Germany
\and
Institut f\"ur Physik, Humboldt-Universit\"at zu Berlin, Germany
\and
LUTH, UMR 8102 du CNRS, Observatoire de Paris, Section de Meudon, France
\and
University of Durham, Department of Physics, U.K.
\and
Laboratoire Leprince-Ringuet, IN2P3/CNRS, Ecole Polytechnique, Palaiseau, France
\and
APC, Paris, France 
\thanks{UMR 7164 (CNRS, Universit\'e Paris VII, CEA, Observatoire de Paris)}
\and
Dublin Institute for Advanced Studies, Ireland
\and
Landessternwarte, K\"onigstuhl, Heidelberg, Germany
\and
Laboratoire de Physique Th\'eorique et Astroparticules, IN2P3/CNRS,
Universit\'e Montpellier II, France
\and
Laboratoire d'Astrophysique de Grenoble, INSU/CNRS, Universit\'e Joseph Fourier, France 
\and
DAPNIA/DSM/CEA, CE Saclay, Gif-sur-Yvette, France
\and
Unit for Space Physics, North-West University, Potchefstroom, South Africa
\and
Laboratoire de Physique Nucl\'eaire et de Hautes Energies, IN2P3/CNRS, Universit\'es
Paris VI \& VII, Paris, France
\and
Institut f\"ur Theoretische Physik, Lehrstuhl IV: Weltraum und
Astrophysik, Ruhr-Universit\"at Bochum, Germany
\and
Institute of Particle and Nuclear Physics, Charles University, Prague, Czech Republic
\and
University of Namibia1, Windhoek, Namibia
\and
European Associated Laboratory for Gamma-Ray Astronomy, jointly
supported by CNRS and MPG}
\offprints{Berrie Giebels \email{berrie@poly.in2p3.fr}}

\abstract{ The High Energy Stereoscopic System (\hess) has observed
  the high-frequency peaked BL Lac object \pks\ in 2003 between
  October 19 and November 26 in Very High Energy (VHE) $\gamma$-rays
  ($E\geq 160\,\rm GeV$ for these observations). Observations were
  carried out simultaneously with the Proportional Counter Array (PCA)
  on board the Rossi X-ray Timing Explorer satellite (RXTE), the
  Robotic Optical Transient Search Experiment (ROTSE) and the
  Nan\c{c}ay decimetric radiotelescope (NRT). Intra-night variability
  is seen in the VHE band, the source being detected with a high
  significance on each night it was observed. Variability is also
  found in the X-ray and optical bands on kilosecond timescales, along
  with flux-dependent spectral changes in the X-rays. A transient
  X-ray event with a 1500 s timescale is detected, making this the
  fastest X-ray flare seen in this object. No correlation can be
  established between the X-ray and the $\gamma$-ray fluxes, or any of
  the other wavebands, over the small range of observed
  variability. The average \hess\ spectrum shows a very soft power law
  shape with a photon index of $3.37 \pm 0.07_{\rm stat} \pm 0.10_{\rm
  sys}$. The energy outputs in the 2--$10\,\rm keV$ and in the VHE
  $\gamma$-ray range are found to be similar, with the X-rays and the
  optical fluxes at a level comparable to some of the lowest
  historical measurements, indicating that \pks\ was in a low or
  quiescent state during the observations. Both a leptonic and a
  hadronic model are used to derive source parameters from these
  observations. These parameters are found to be sensitive to the
  model of Extragalactic Background Light (EBL) that attenuates the
  VHE signal at this source's redshift ($z=0.117$).


  }

\maketitle

\section{Introduction}

The innermost regions of active galactic nuclei, where the largest part
of their luminosity is emitted, can be probed through observations of
their flux variability at different wavelengths. The physical
processes in their central engines and jets are usually considered the main
candidates for the origin of the observed variability. Measurements of
correlated variability, spectral variations and time lags across the
broad-band observations allow modelling of particle distributions and
their radiation processes, as well as probing the acceleration
mechanisms that are involved.

\pks\ is probably the most prominent and best-studied blazar-type
Active Galactic Nuclei (AGN) in the Southern Hemisphere. The emission
of \pks, its possible variability patterns, as well as correlations
across all wavebands, have been studied exhaustively over the past 20
years (see e.g. \cite{urry}). Its first detection at VHE $\gamma$-rays
by the Durham Mk VI telescopes (\cite{cha99}) classified it as a TeV
blazar, like the northern hemisphere BL Lac objects Mkn~421, Mkn~501,
H~1426$+$428, or 1ES~1959$+$650. Its redshift of $z=0.117$ makes it
the second most distant confirmed TeV blazar after H~1426$+$428
($z=0.129$). \pks\ was the brightest BL Lac object in the EUVE all-sky
survey (\cite{marsh}). This source was confirmed as a high energy
$\gamma$-ray emitter by \hess\ (\cite{aha03a}, AH04 hereafter) at the
$45\sigma$ significance level, when strong detections were reported
for each of the dark periods of observations.

Here we report on simultaneous \hess\ VHE $\gamma$-ray, RXTE/PCA
X-ray, ROTSE optical, and NRT decimetric observations of \pks\ during
the dark periods of October and November 2003. No simultaneous
multi-wavelength campaign had before included an Atmospheric Cherenkov
Telescope (ACT) that could sample the evolution of the high energy
component of the spectral energy distribution (SED) of this object. We
also include EGRET archival data, and other archival radio through
X-ray data obtained from the NASA/IPAC Extragalactic Database
(NED). Details of the observations and data reduction/analysis are
given in \S \ref{sect:obs}. Light curves and spectra are described in
\S \ref{sect:res}. The attenuation of the \hess\ spectrum by the EBL
and an interpretation of the data using a leptonic and a hadronic
model are discussed in \S \ref{sect:dis}.

\section{Observations and data analysis}  \label{sect:obs}

\subsection{H.E.S.S.}

\subsubsection{H.E.S.S. detector}
In its first phase, the \hess\ array consists of four atmospheric
Cherenkov telescopes operating in stereoscopic mode. However, the data shown
here were taken during the construction of the system, initially with
two telescopes, with one more as of mid-September, 2003. The fourth
and final telescope was added to the array in December 2003,
subsequent to the data presented here. Each telescope has a
tessellated $13\,\rm m$-diameter ($107\,\rm m^2$ surface area) mirror which
focuses the Cherenkov light from the showers of secondary particles
created by the interaction of $\gamma$-rays in the atmosphere onto a
camera in the focal plane. This camera consists of 960
photomultipliers, each with a pixel size of $0.16^\circ$, giving a
field of view of $5^\circ$. For the data sets presented here the
$\gamma$-ray trigger threshold is $\approx 100\,\rm GeV$ and the
spectral threshold is $160\,\rm GeV$ with an energy resolution $\simeq
15\%$. The experiment is located in the Khomas highlands in Namibia,
($23^\circ\,\rm S$, $15^\circ\,\rm E$, $1800\,\rm m$ a.s.l.). A
detailed description of the layout and the components of the telescope
optical systems, including the segmented mirror with its support
structure, the mirror facets for each telescope and the Winston cone
light concentrators in front of the PMT camera can be found in
\cite{bern}, and a description of the mirror alignment is given in
\cite{cornils}. For details on the camera calibration see
\cite{bruno}. The trigger system is described in \cite{funk}. Early
reports of \hess\ have been given elsewhere (see e.g. \cite{hof}).

\begin{figure}[htb!]	       	
\begin{center}	
\includegraphics[angle=0,scale=0.48]{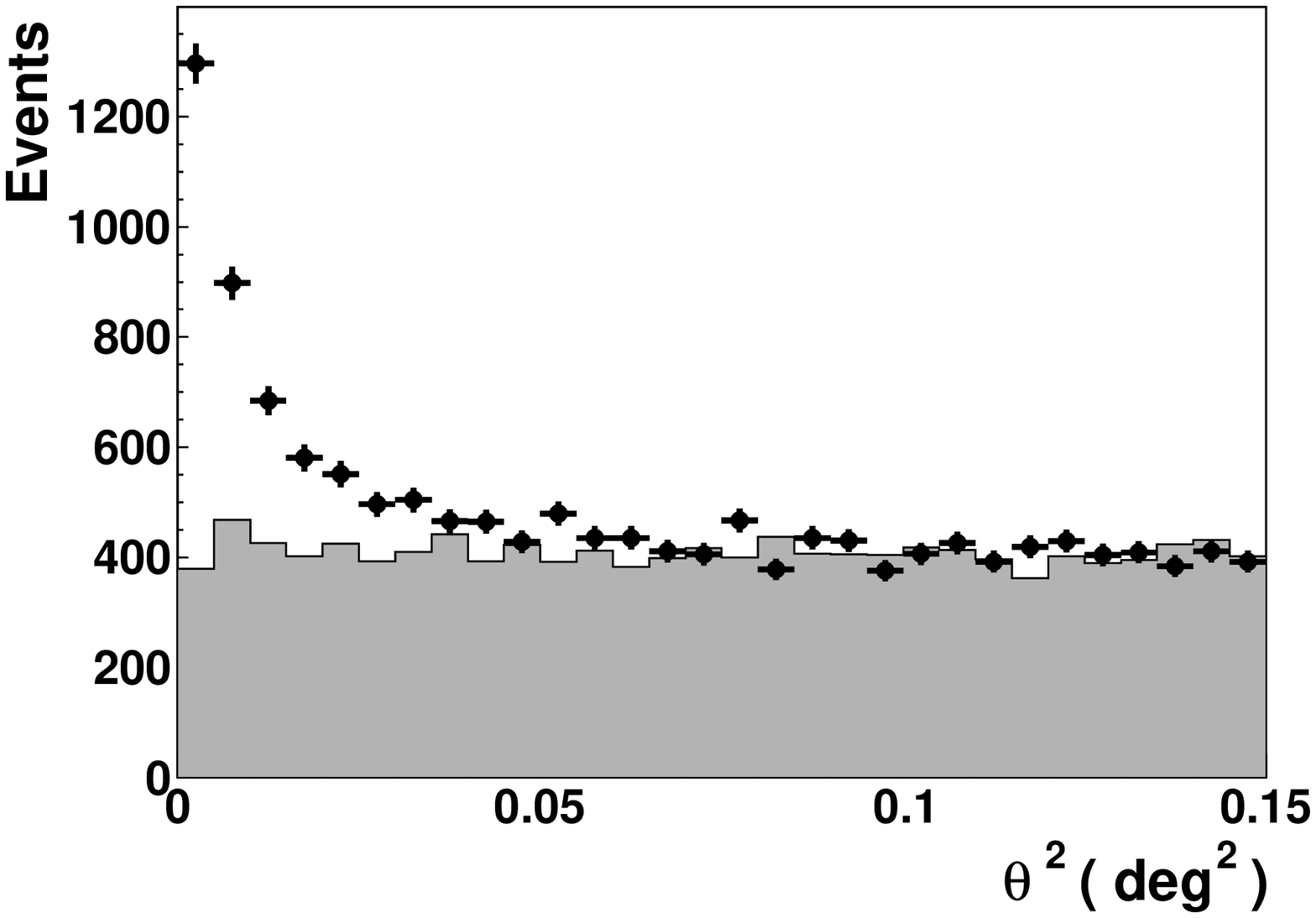} 
\includegraphics[angle=0,scale=0.48]{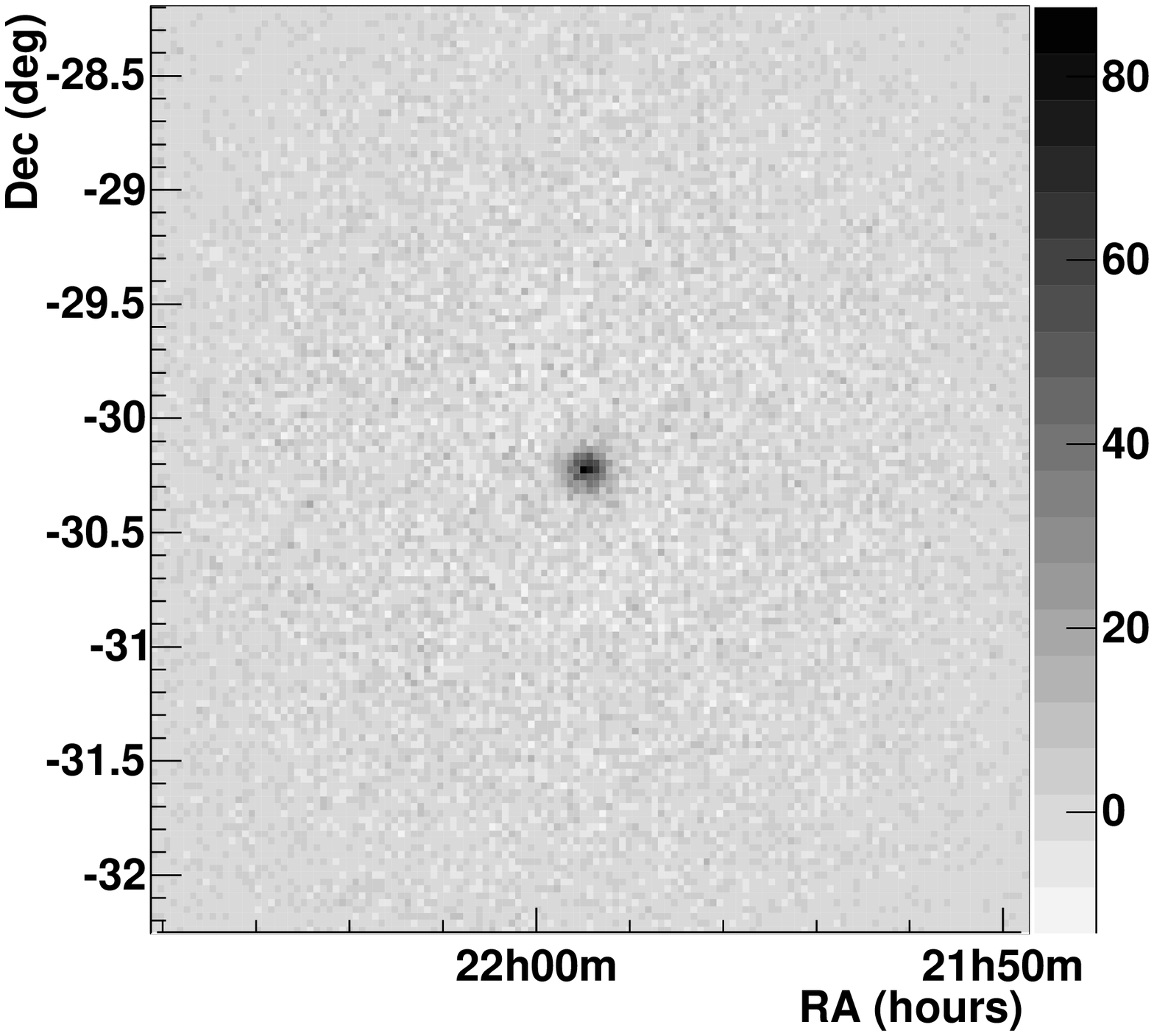}
\end{center}
\caption{Top: The distribution of $\theta^2$ for on-source events
  (points) and normalized off-source events (shaded) for the
  observations in October and November 2003. Bottom: The
  two-dimensional distribution of excess events observed in the
  direction of PKS 2155-304. The bins are not correlated and represent
  the actual distribution of observed $\gamma$-rays on the sky. The
  right hand scale is the number of counts.}
\label{fig:f1}
\end{figure}

\subsubsection{Observations}
In the beginning of the observation period of October 2003, a single 5~$\sigma$
detection by \hess\, achieved in 1h, triggered an approved \xte\ ToO (Target of
Opportunity) on this target. The results presented here are based on
observations carried out in 2003 between October 19 and November 26.

The data were taken in the \textit{Wobble} mode where the source
direction is positioned $\pm 0.5^\circ$ in declination relative to the
centre of the field of view of the camera during observations. This
allows for both on-source observations and simultaneous estimation of
the background induced by charged cosmic rays. The data reported here
are selected and analyzed with the ``standard analysis'' described in
\S 4 of AH04. The background is estimated here by using all events
passing cuts in a ring around the source, as described in section 4.3
in AH04. The runs passing the quality selection criteria total 32.4 hours of
livetime on the source. The total two-dimensional significance sky map
is shown in Figure \ref{fig:f1}, along with a graph showing the
$\theta^2$ distribution (where $\theta^2$ is the square of the angular
difference between the reconstructed shower position and the source
position) of the 1764 excess events observed. This yields a detection
at the $34.3\sigma$ level, at the
average rate of $0.91\gamma/\,\rm min$ and a significance of
$6.0\sigma/\sqrt{\mathrm{h}}$.

The methods used here for reconstructing the energy of each
event and for determining a spectrum are described in \S 6 of AH04.
The measured time-average spectrum is fitted by a power law of the
form ${\rm d}N/{\rm d}E = I_0 (E/1~{\rm TeV})^{-\Gamma}$ with $I_0$
the flux normalization at $1\,\rm TeV$ and $\Gamma$ the photon index.
The photon index obtained from the time-averaged spectrum is then used
as a fixed parameter to estimate the integral flux above $300\,\rm
GeV$ for each run. This integrated flux takes into account the
effective area and threshold variations due to the source moving
through the sky, giving more reliable variability information than
counting rates in units of $\gamma$-rays/minute. Overall systematic errors
are estimated to be 20\% for the integral flux and $\simeq0.1$ for the
photon index.

\subsection{RXTE}

The PCA (\cite{jah}) units of RXTE observed \pks\ between October 22
and November 23 of 2003 with exposures of typically $10\,\rm ks$ in
October and $\simeq 1\,\rm ks$ in November. The STANDARD2 data were
extracted using the ftools in the HEASOFT 5.3.1 analysis software
package provided by NASA/GSFC and filtered using the RXTE Guest
Observer Facility (GOF) recommended criteria. The changing
Proportional Counter Units (PCU0/2/3) configuration throughout the
observations was taken into account in the data reduction, and only
the signals from the top layer (X1L and X1R) were used. When reducing
the PCUs individually to establish the time-averaged spectrum, the
average spectral fit parameters are similar within error bars but with
a systematically higher $\chi^2$ for spectral fits performed on PCU0
data alone. On shorter timescales this effect becomes negligible and
PCU0 contributes to the statistical significance of the flux
measurement. Therefore all PCUs were kept for the overall light curve
which is binned in $400\,\rm s$ bins, but only PCU2 and PCU3 are used
for the analysis of data segments that are simultaneous with \hess\
runs and for the time-averaged spectrum.

The average spectrum used in the SED is derived by combining PCU2 and
PCU3 spectra using the {\tt addspec} tool weighted by the counts
information delivered by {\tt fstatistic} and then the corresponding
response matrices were combined with {\tt addrmf}. The
faint-background model was used and only the 3--40 PHA channel range
was kept in {\tt XSPEC v. 11.3.1}, or approximately 2--$20\,\rm keV$.

To build a light curve in units of integrated flux in the 2--$10\,\rm
keV$ band, spectral data were derived from $400\,\rm s$ bins to probe
short timescales with adequate statistical accuracy. These segments
are then fitted by a power law in XSPEC with PCU
configuration-dependent response matrices generated by the ftool {\tt
pcarsp v. 10.1} and a fixed column density of $N_{\rm H} = 1.7\times
10^{20}\,\rm cm^{-2}$ obtained from PIMMS\footnote{See
http://legacy.gsfc.nasa.gov/Tools/w3pimms.html}. This yields the flux
and the error (corresponding to the $1\sigma$ confidence level) on the
flux reported in the light curves in Figure \ref{fig:f2}, in units of
$10^{-11}\,\rm erg\,cm^{-2}\,s^{-1}$ in the 2--$10\,\rm keV$ band. The
fits did not improve by using a broken power-law for the $400\,\rm s$
binned observations.

\subsection{ROTSE}  \label{ssect:rotse}

The ROTSE-III array is a worldwide network of four $0.45\,\rm m$
robotic, automated telescopes built for fast ($\approx 6\,\rm s$)
response to GRB triggers from satellites such as HETE-2 (High Energy
Transient Explorer 2) and Swift. The ROTSE-III telescopes have a wide
($1.85^\circ\times1.85^\circ$) field of view imaged onto a Marconi
$2048\times 2048\,\rm pixel$ back-illuminated thinned CCD and
 are operated without filters. The ROTSE-III systems are described in detail
in \cite{2003PASP..115..132A}. At the time of the observations of
\pks\ in October and November 2003, two ROTSE-III telescopes were
operational in the Southern hemisphere: ROTSE-IIIa located at the
Siding Spring Observatory, Australia and ROTSE-IIIc at the
H.E.S.S. site. The ROTSE-IIIc telescope is located in the centre of
the H.E.S.S. telescope array. A 30\% share of the total observation
time is available to the H.E.S.S. collaboration, which has been used
to perform an automated monitoring programme of blazars, including
objects that are being observed with the H.E.S.S. telescopes. Both
telescopes participated in the observation campaign on \pks\ in
October and November 2003.

The telescopes observed \pks\ typically 10 times per night
taking sequences of 2 frames with $60\,\rm s$ exposures with a slight
dithering of the pointing to reduce the impact of individual noisy
pixels. The typical limiting magnitude, depending on the sky conditions,
is $18.5^\mathrm{mag}$. Overall, 323 bias-subtracted and flat-fielded
frames have passed visual inspection and are used to produce a light
curve. A total of 6 frames were rejected due to the presence of stray
light from Jupiter.

Using an overlay of 50 isolated comparison stars with similar
brightness ($12^\mathrm{mag}-14^\mathrm{mag}$) and co-located with
\pks\ ($<15\,{\rm arcmin}$), a two-dimensional Gaussian is fit to
the intensity distribution characterising the point-spread function by
$\sigma_\mathrm{psf}$.

To estimate the local sky-background for the reference stars and the
target object, an annulus with inner radius $2\times
\sigma_\mathrm{psf}$ and an outer radius $6\times \sigma_\mathrm{psf}$
is chosen. Based upon a reference frame which is derived from
co-adding 30 individual frames, a mask is calculated for each object
excluding regions where faint objects coincide with the annulus;
pixels exceeding 3 standard deviations of the local sky background are
excluded. Using the local sky background, the intensity and error of
each object is calculated. Using the 50 reference stars, a relative
intensity and statistical error with respect to a reference frame is
calculated.

 The absolute flux values are obtained by calculating a relative $R$
magnitude by comparing the instrumental magnitude with the USNO
catalogue as described in \cite{akerl02}. The procedure has been
checked by comparing the average $R$ magnitude of a sample of 70 BL
Lac type objects determined with ROTSE observations carried out over
one year of operation with the $V$ magnitude listed in the $10^{\rm
th}$ Veron Cetty \& Veron catalogue of BL Lac type objects. The
average $V-R$ of 0.5 that is found is consistent with the average
value for $V-R$ obtained from cross-checking the colours with the
2MASS catalogued value for the BL Lac type objects.

 Finally, the host galaxy has been resolved in optical (\cite{fal})
and NIR (\cite{1998A&A...336..479K}, KFS98 hereafter) and found to be
a typical giant elliptical with ${\rm M}(R)=-24.4$ which translates
into an apparent ${\rm m}(R)=15.1$ (here the distance moduli given by
KFS98 have been used to calculate the apparent magnitude based upon
the absolute magnitude quoted). The ROTSE measurements have as maximum
and minimum ${\rm m}(R,{\rm min})=13.3$ and ${\rm m}(R,{\rm
max})=13.7$ which corresponds to
$10\,\rm mJy$ for the maximum observed flux and
$6.7\,\rm mJy$ for the minimum flux taking the contribution of the host
galaxy into account. These values are considerably lower than the
retrieved archival data indicating that \pks\
was in a low state at the moment of the observations.

\subsection{NRT}  \label{ssect:nancay}
The Nan\c{c}ay radiotelescope is a single-dish antenna with a
collecting area of $200\times34.56$ m$^2$ equivalent
to that of a $94\,{\rm m}$-diameter parabolic dish (\cite{vandriel}). The
half-power beam width at $11\,\rm cm$ is $1.9\,{\rm arcmin}$ (EW)
$\times 11.5\,{\rm arcmin}$ (NS) (at zero declination), and the system
temperature is about $45\,\rm K$ in both horizontal and vertical
polarizations. The point source efficiency is $0.8\,{\rm K}\,{\rm Jy}^{-1}$, and the
chosen filter bandwidth was 12.5 MHz for each polarization, split
into two sub-bands of 6.25 MHz each. Between 4 and 14 individual
1-minute drift scans were performed for each observation, and the flux
was calibrated using a calibrated noise diode emission for
each drift scan. Data processing has been done with the Nan\c{c}ay
local software NAPS and SIR.

A monitoring programme with this telescope on extragalactic sources
visible by both the NRT and \hess\ is in place since 2001. For the
campaign described here it consisted of a measurement at $11\,\rm cm$
every two or three days. The average flux for the 8 measurements in
October and November 2003 was $0.30\pm0.01\,\rm Jy$ with possible
marginal variability.

\section{Results} \label{sect:res}
\subsection{Light curves}

\begin{figure*}[htbp]
{\includegraphics[angle=0,scale=1]{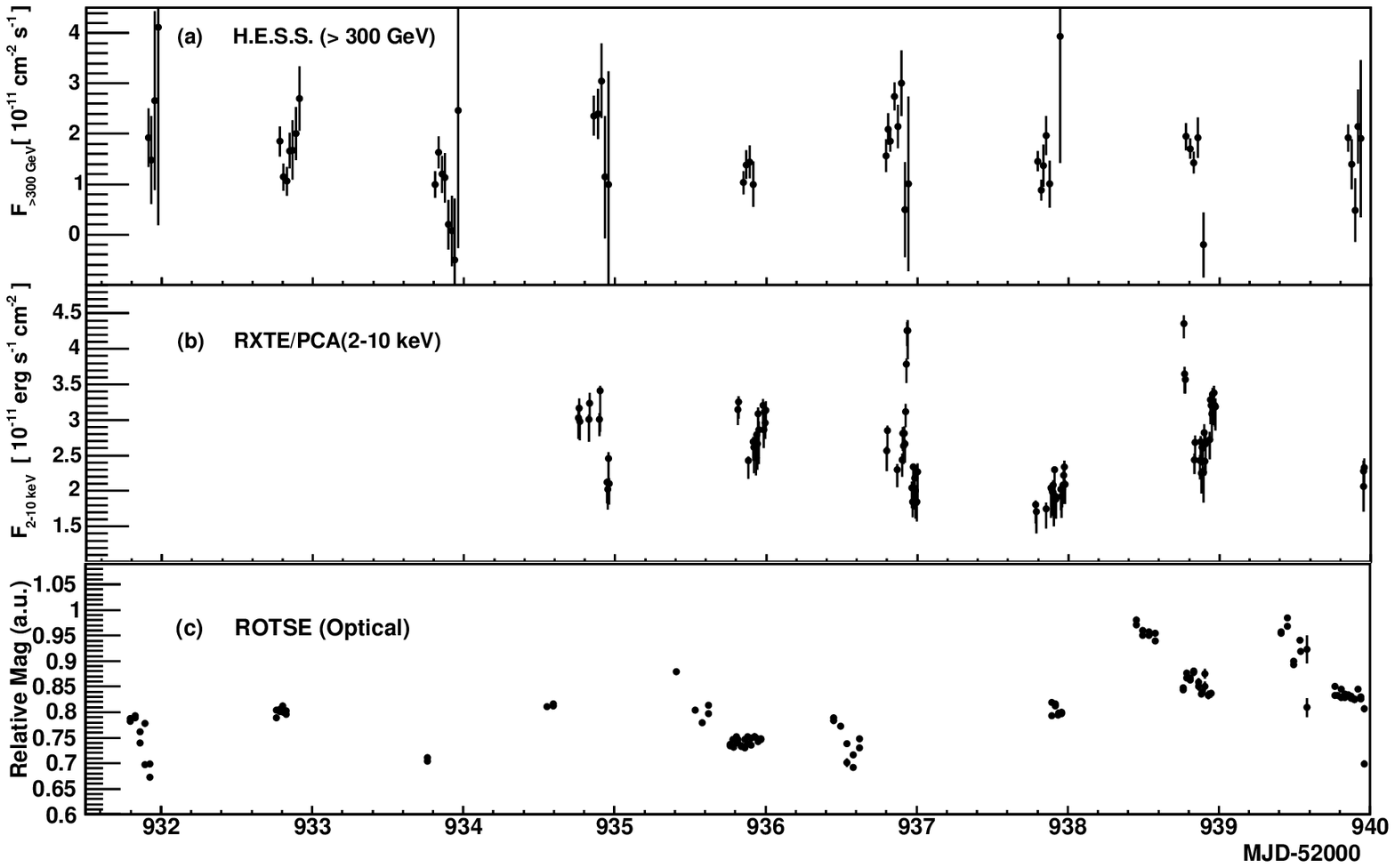}}
{\includegraphics[angle=0,scale=1]{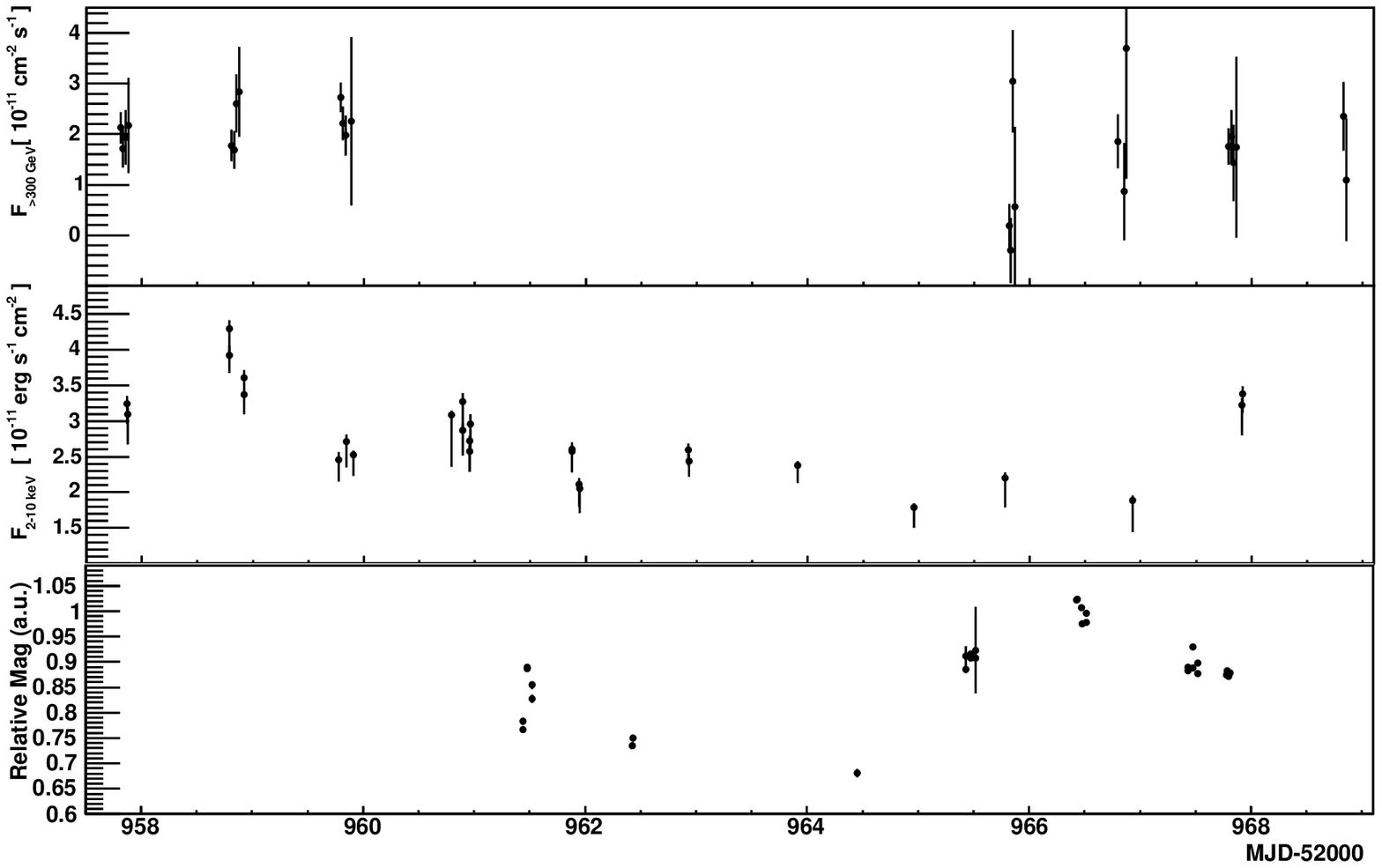}}
\caption{Top: (a) October 2003 Light curve from H.E.S.S. binned in run
  lengths averaging 28 minutes each, in units of integral flux above
  300 GeV. (b) Light curve expressed as flux in the 2--$10\,\rm keV$
  band for the \xte\ observations, in $400\,\rm s$ bins. (c) Light
  curve derived from the ROTSE optical data. Bottom: Same as above but
  for the November 2003 data. Note that the \xte\ observations were
  much shorter than in the previous month.}
\label{fig:f2}
\end{figure*}

\begin{figure}[htb!]
{\includegraphics[angle=0,scale=0.5]{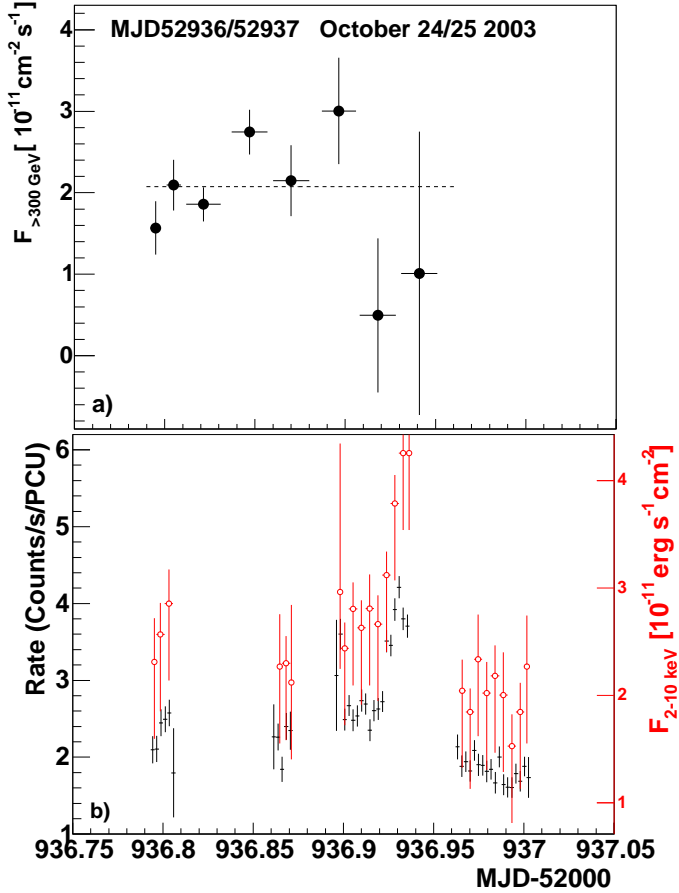}}
\caption{(a) The VHE light curve derived run by run on MJD52936. The
    horizontal error bars are the length of the run from which the
    flux is derived. The dashed line is the result of a $\chi^2$ fit
    of the data to a constant, which yields a $\chi^2$ of 15 for 7
    degrees of freedom, corresponding to a 0.04 $\chi^2$ probability
    which is evidence for the variability seen here. (b): The
    2--$10\,\rm keV$ X-ray flux (\textit{open circles, right scale})
    and counting rate normalized to 1 PCU (\textit{lines, left scale})
    showing the fast transient.}
\label{fig:flare}
\end{figure}

The October and November 2003 light curves of all the \hess, \xte\ and
ROTSE observations are shown in Figure \ref{fig:f2}. The \hess\ light
curve is binned in run-length times averaging 28 minutes each. The
flux is in units of $10^{-11}\,\rm photons\,cm^{-2}s^{-1}$ above
$300\,\rm GeV$, derived using the average photon index 3.37 obtained
in section \ref{ssect:spect}. Spectra could not be derived on a
run-by-run basis due to the weak signal. As for the observations reported in AH04, the overall
light curve is inconsistent with a constant flux. A $\chi^2$ fit of
the data to a constant yields a $3.4\times 10^{-10}$ $\chi^2$
probablility. The intra-night VHE flux on MJD52936 (Fig. \ref{fig:flare})
exhibits an increase of a factor of $1.9\pm0.6$ in $0.11\,\rm d$. On
MJD52932 the peak-to-peak flux shows an increase of a factor of
$2.5\pm0.9$ within $0.09\,\rm d$. These timescales are longer than the
30 minute doubling time reported in AH04. For these two extreme cases of
VHE variability observed during this campaign only the second had a
limited \xte\ coverage.

The 2--$10\,\rm keV$ X-ray flux in this campaign ranges from
$F_{\mathrm{2-10~keV}} = 2.0\times 10^{-11}\,\rm erg\,cm^{-2}\,s^{-1}$
to $4.4\times 10^{-11}\,\rm erg\,cm^{-2}\,s^{-1}$. The maximum is lower
than the 20 November 1997 measurement of $2.3\times 10^{-10}\,\rm
erg\,cm^{-2}\,s^{-1}$ (\cite{vest}) indicating that the X-ray state seen
here is not exceptionally high. The minimum seen here is consistent
with historically low fluxes (\cite{zhang}). The intra-night
variability is also obvious here, but no flare was completely resolved.

 A 60\% flux variability in $t_{\rm var}\approx1.5\,\rm ks$ on MJD52936 is
the best marked transient episode in the observations reported here
(bottom panel b) in Figure \ref{fig:flare} for which the \hess\
observations were made at the end of the transit inducing a large
associated error on the flux estimation due to the high zenith angle
of the source. This timescale is comparable to those reported by
\cite{gaidos} where doubling times as short as 15 min from Mkn~421
were observed in the VHE band.

This flare is the fastest rise seen in this object to date since BeppoSAX
saw a $5\times10^4\, \rm s$ rise timescale (\cite{zhang}) and \cite{kata}
observed a doubling timescale of $3\times10^4\,\rm s$ with the ASCA
satellite. So far the fastest rise in this type of object was observed
in Mkn~501 with a 60\% increase in less than 200 s (\cite{cata})
though \cite{xue} claims that this flare is likely to be an
artifact.

The optical emission of \pks\ is dominated by the nucleus which
outshines the host galaxy by a factor of $\approx 4$
given in KFS98. The observed variability amplitude is
therefore not biased by the constant emission of the host galaxy and
mainly due to the activity of the nucleus. The peak-to-peak amplitude of
variability is moderate compared to the variability amplitude at shorter
wavelengths and typically $0.1^\mathrm{mag}$ peak-to-peak.
The object has been monitored over longer time-scales with the ROTSE-IIIc telescopes showing variations with amplitudes close to
$1^\mathrm{mag}$.

\subsection{Correlation analysis}
\begin{figure}[htb!]
{\includegraphics[angle=0,scale=0.65]{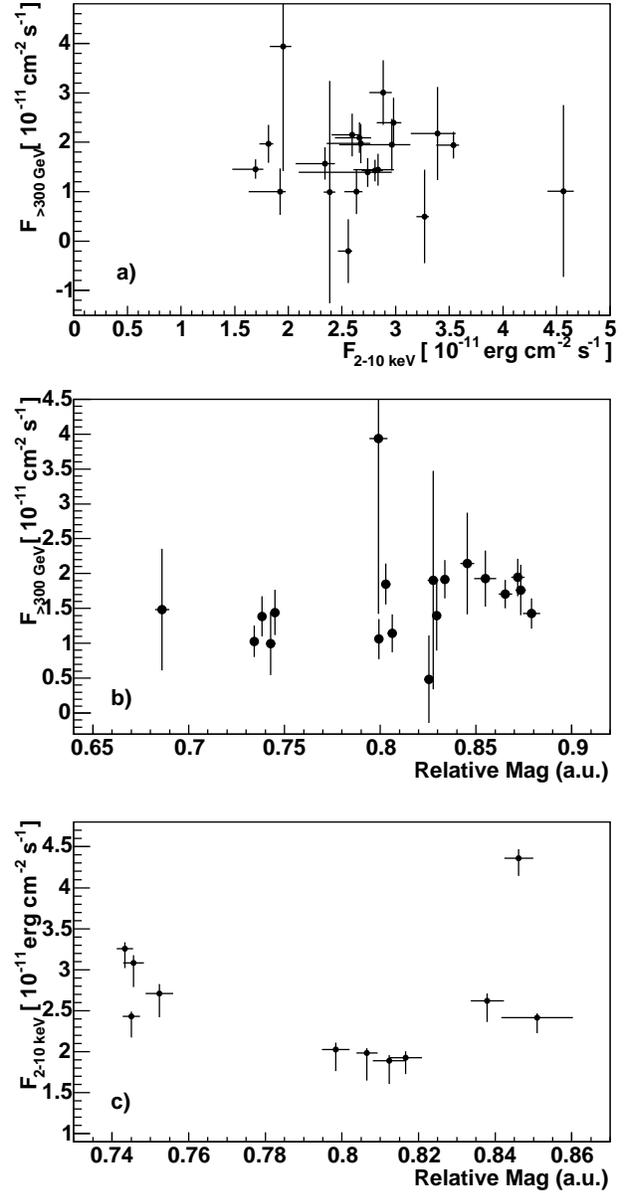}}
\caption{Correlation plots between different wavebands. No clear
  correlation is found between any of the simultaneous measurements. (a) Correlation
  plot for the 23 X-ray data segments that overlapped exactly with a
  \hess\ observation. (b) Simultaneous \hess\ and
  ROTSE observations. (c) Simultaneous \xte\ and ROTSE
  observations. The optical data were binned to the 400 s long RXTE
  segments, which are overlapping on 6 different days.}
\label{fig:f3}
\end{figure}

\begin{figure*}
{\includegraphics[angle=0,scale=0.8]{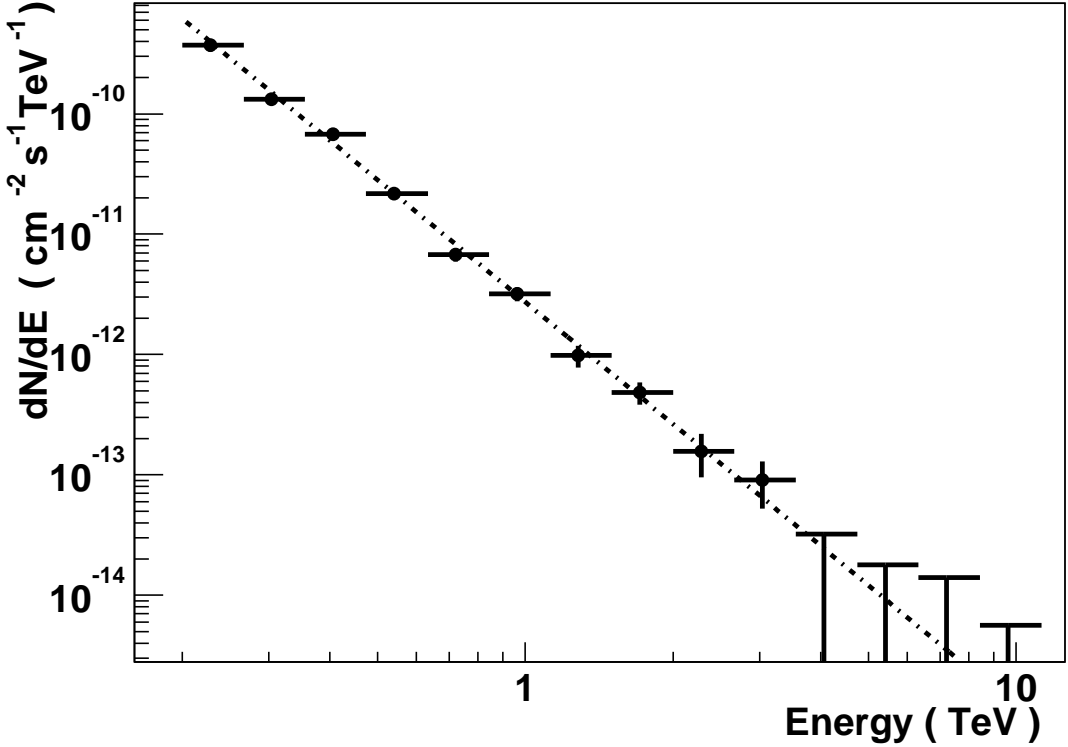}}
\caption{Time-averaged spectrum derived from the October and November
  2003 \hess\ data along with a fit to a powerlaw.}
\label{fig:f5}
\end{figure*}

\begin{figure*}

{\includegraphics[angle=270,scale=0.6]{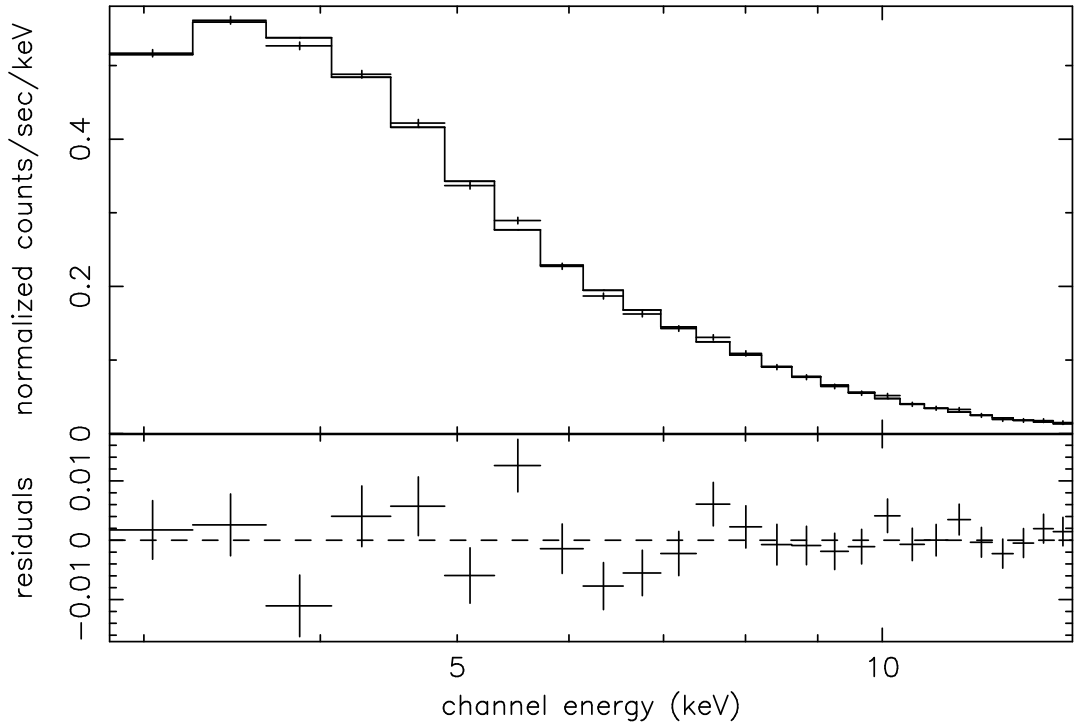}}
\caption{X-ray spectrum derived from the summed October and November 2003 data,
  using PCU2 and PCU3, fitted by a broken power law. The top panel
  shows the data and folded model, the bottom panel shows the
  residuals between the data and the model.}
\label{fig:xspec1}
\end{figure*}

In order to quantitatively look for correlated variability between the
VHE, X-ray and optical bands, the measured fluxes are plotted against
each other in Figure \ref{fig:f3} for all the observations carried out
during this campaign. For correlated VHE/X-ray variability, the \xte\
analysis was slightly modified: only observation segments that happen
exactly within a \hess\ run are reduced and analyzed (using only PCU2,
PCU3 and combinations thereof). This provides 23 simultaneous data
segments for the whole campaign for which the fluxes are represented
in Figure \ref{fig:f3}. There is no obvious correlation for those
observations (correlation factor $r=0.37\pm0.13$). For correlated
optical/VHE variability, ROTSE observations that happen within a
\hess\ run are averaged and their errors summed quadratically. No correlation ($r=0.24\pm0.27)$ is found for these
observations. Also no correlation was found between the optical and
X-ray band ($r=-0.02\pm0.05)$.

\subsection{Spectra}
\label{ssect:spect}

The October and November 2003 \hess\ data were all combined for the
spectrum shown in Figure \ref{fig:f5}. The best-fitting power law
($\chi^2=7$ for 8 degrees of freedom) is given by:
\begin{center}
\begin{multline} \frac{{\rm d}N}{{\rm d}E}=(2.73 \pm 0.17) \times
  10^{-12}\left( \frac{E}{\mathrm{TeV}}
  \right)^{-3.37\pm0.07\pm0.10} \\ \mathrm{cm}^{-2}\,\mathrm{s}^{-1}\,\mathrm{TeV}^{-1}
\label{eq:eq1}
\end{multline}
\end{center}
which is comparable to $\Gamma=3.32\pm0.06$ and $I_0=(1.96 \pm 0.12)
  \times 10^{-12}\,\mathrm{cm}^{-2}\,\mathrm{s}^{-1}\,\mathrm{TeV}^{-1}$
  previously reported in AH04.

The result of the broken power law fit for the combined \xte\ PCU2 and
PCU3 spectrum is shown in Figure \ref{fig:xspec1}. It yields an
unabsorbed flux in the 2--$10\,\rm keV$ band of $F_{\mathrm{2-10~keV}}=(2.66\pm0.04)\times 10^{-11}\,\rm erg\,cm^{-2}\,s^{-1}$ 
($\chi^2 = 41$, 31 degrees of freedom), a lower index of
$\Gamma_L=2.81\pm0.05$, a break energy of $E_{b}=4.9\pm0.8\,\rm keV$ and a
higher index of $\Gamma_H=2.95\pm0.04$. A single power law fit to the
same data yields 
$F_{\mathrm{2-10~keV}}=(2.69\pm0.03) \times 10^{-11}\,\rm erg\,cm^{-2}\,s^{-1}$ 
($\chi^2 = 51$, 34 degrees of freedom) and
an index of $\Gamma=2.88\pm0.13$, a poorer fit than the broken power
law, but the index still provides information that can be used for
comparison with historical measurements. Indeed, the derived index is
close to those measured by the BeppoSAX satellite (\cite{giommi}),
GINGA (\cite{sembay}) and well within the range observed by
EXOSAT (\cite{treves}). The statistics above $10\,\rm keV$ in our
\xte\ observations are too poor to check the existence of a possible hard tail
above $20\,\rm keV$ (\cite{giommi}) which might be
the signature of the onset of a high-energy component.

\begin{figure}
{\includegraphics[angle=0,scale=0.50]{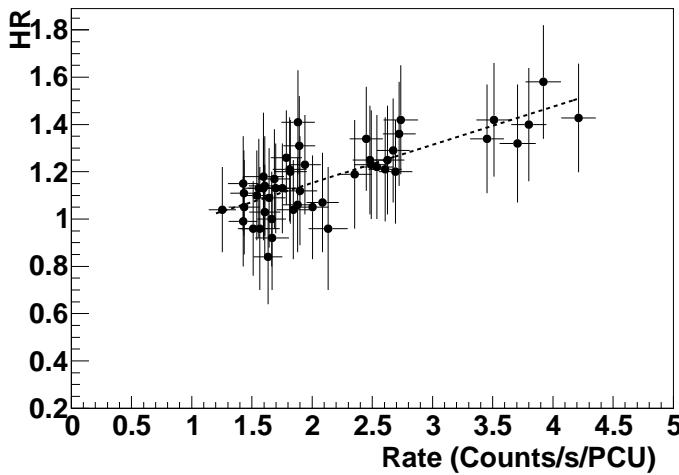}}
\caption{Plot of the hardness ratio HR versus the counting rate normalized to 1 PCU.}
\label{fig:f4}
\end{figure}

In order to look for flux dependent spectral variability, the \xte\
data subset used in Figure \ref{fig:flare} is divided into two energy
bands, the PHA channels 0--9 (soft band) and 10--27 (hard band),
corresponding to approximately 1--4 keV and 4--11 keV, respectively. A
hardness ratio (HR), shown in Figure \ref{fig:f4}, is the ratio of the
counting rate in the hard band over the soft band. There is a clear
correlation of the HR with the rate, peaking when the rate is
highest. The correlation factor between the rate and the HR is
$r=0.76\pm0.12$. Even though the variability timescale here is much
smaller, this behavior is compatible with the hardening reported in
\cite{chia}.

\subsection{EBL corrected spectrum}
\label{ebl}
 \begin{figure}[htb!]
 {\includegraphics[angle=270, scale=0.37]{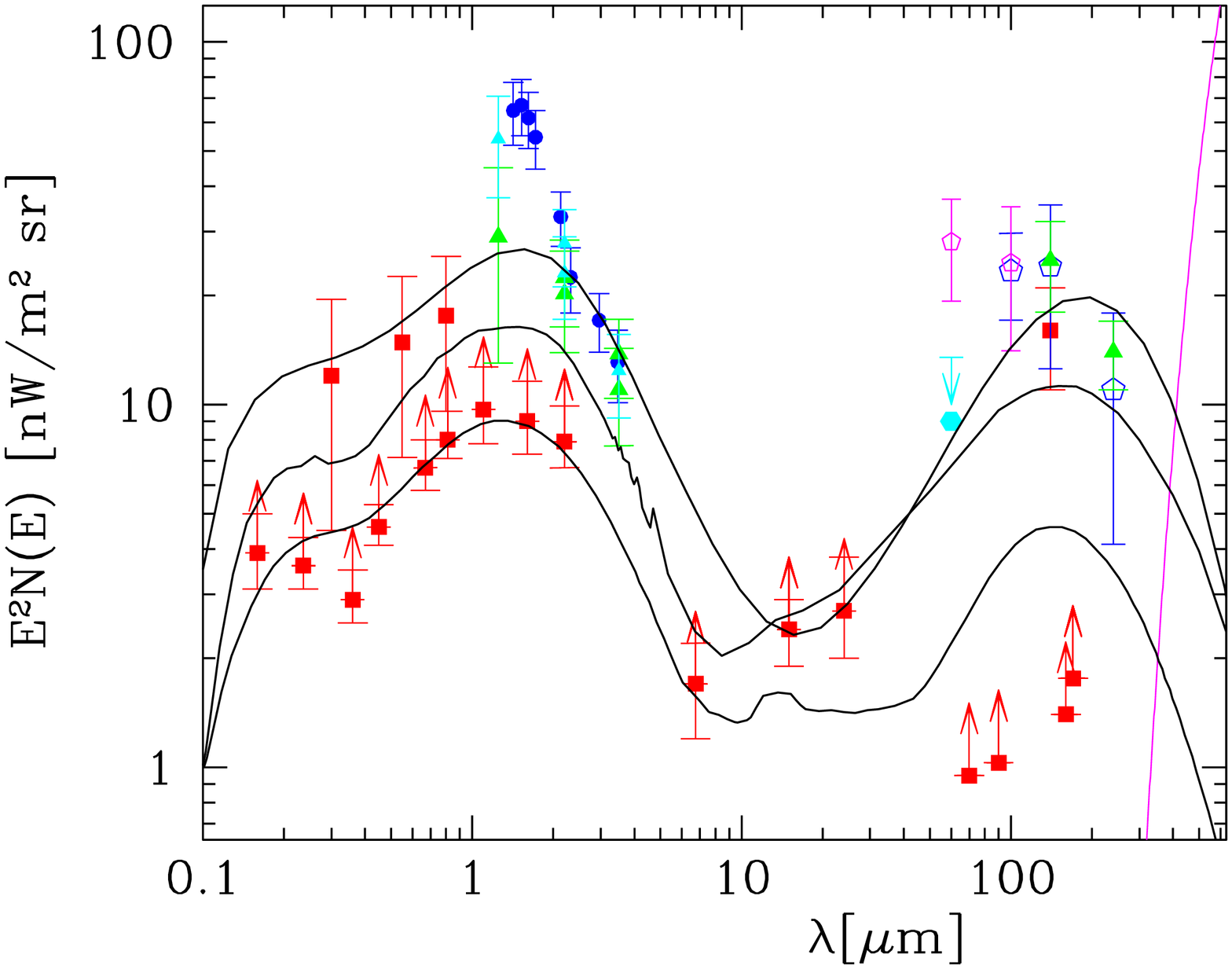}}
 \caption{Spectral energy distribution of the Extragalactic Background
Light (EBL). It is thought to be characterized by two distinct bumps,
around 1--2 and 100--$200\, \mu\rm m$, produced by the red-shifted
stellar light and re-radiated thermal dust emission, respectively. The
compilation of measurements have been taken from \cite{haus} and
\cite{1426}. Lower limits are from HST and ISO source counts
(\cite{madau,elbaz,gardner}). Above $400\, \mu\rm m$, the contribution
of the CMB starts to dominate. The three continuous lines represent
the ``Primack-type'' EBL shapes: from higher to lower fluxes (at
$1\,\mu \rm m$), Phigh, Primack01 and Primack04 models (see text).
The steep line starting at $300\, \mu\rm m$ is the onset of the CMBR. }
 \label{fig:ebl1}
 \end{figure}

For objects at non-negligible redshifts, the large, energy-dependent
opacities can cause the emitted spectrum to be greatly modified both
in shape and intensity (see e.g. \cite{steck,biller,coppi,vassiliev}).
Unfortunately, at present the knowledge of the EBL still has large
uncertainties, for both direct measurements and models, as summarized
in \cite{primack}. In order to estimate the intrinsic VHE spectrum, and
thus to locate the Inverse Compton (IC) peak of the blazar's SED, we
have used three EBL models (Fig. \ref{fig:ebl1}) as representatives of
three different flux levels for the stellar peak component. This is
the EBL energy range which mostly affects the \hess\ spectrum: with
data up to 3--$3.5\,\rm TeV$, the peak of the \gamgam cross section is
reached for soft photons with wavelengths $\lesssim 5\, \mu\rm m$.

 \begin{figure*}
{\includegraphics[scale=0.6,angle=270]{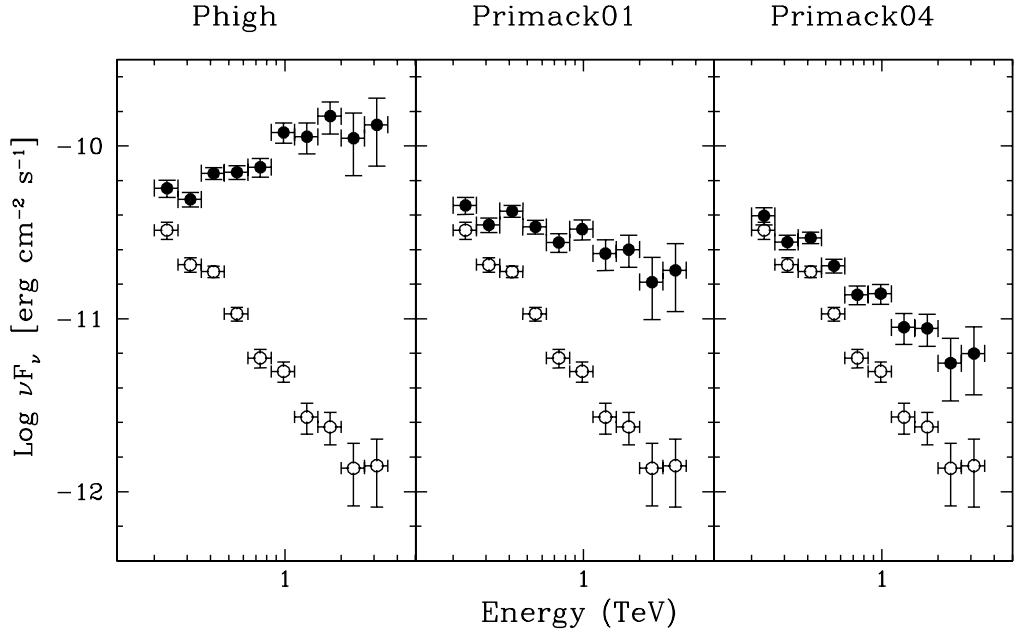}}
\vspace{-8ex}
 \caption{\pks\ absorption-corrected spectra, along with the observed
spectrum, for each EBL model considered and in a $\nu F_\nu$ plot (i.e. a
zoom in the blazar's SED). Open dots: \hess\ observed
spectrum. Filled dots: reconstructed intrinsic spectrum. All spectra
from ``Primack-type'' models are compatible with a power-law shape,
with different slopes: hard (Phigh) and soft (Primack01, Primack04),
locating the intrinsic SED peak for the high component above 1--$2\,\rm TeV$
or below $200\,\rm GeV$, respectively.}
 \label{fig:ebl3}
 \end{figure*}

The three models used here are (in order from higher to lower fluxes):
the phenomenological shape used in \cite{1426}, which is based on the
original \cite{primack} calculation but smoothed and scaled up to
match the data points below $1\, \mu\rm m$ and at 2--$3.5\, \mu\rm m$
(hereafter {\it Phigh}); the original \cite{primack} calculation for a
Salpeter initial mass function (hereafter Primack01); and the new 2004
calculation (\cite{primack04}, hereafter Primack04), which takes
advantage of the recent improvements in the knowledge of the
cosmological parameters and of the local luminosity function of
galaxies.

The opacities are calculated from the EBL SED shapes taking into
account only the cosmology ($H_0=70\,\rm km/s/Mpc$, $\Omega_{\rm
Mat}=0.3$ and $\Omega_{\Lambda}=0.7$). To treat all three shapes
similarly, no evolution has been introduced at this point. This
corresponds to a ``maximum absorption" hypothesis (i.e., for
increasing $z$, constant instead of decreasing EBL comoving energy
density). But at these redshifts ($\sim0.1$) and, for example,
assuming the evolution given in \cite{primack}, the difference is
still small (variation in the photon index $\Delta\Gamma<0.1$ in the
range 0.3--$1\,\rm TeV$), and negligible compared to the differences
between models.

The resulting absorption-corrected spectra are shown in
Figure \ref{fig:ebl3}, together with the observed spectrum. The
intrinsic spectra are all well fitted by a single power-law model,
with a hard spectrum for Phigh ($\Gamma\approx1.5$), and soft spectra
for Primack01 and Primack04 models ($\Gamma\approx2.3$ and
$\Gamma\approx2.8$, respectively). This effect is directly related to
the different flux levels of the stellar peak component, which imprint
a different amount of softening onto the original spectrum. This
direct link thus yields two simple scenarios for the location of the
blazar's high energy peak, with the dividing line represented by the
EBL flux which gives $\Gamma=2.0$ (1.3 times the Primack01 model).
Models with stellar peak fluxes above this (such as Phigh, and
generally all those in agreement with the direct estimates of the
fluxes between 2 and $3.5\, \mu\rm m$) imply a hard intrinsic spectrum
($\Gamma<2$), and thus an IC peak above 1--$2\,\rm TeV$. EBL models
with lower fluxes (such as the Primack01 and Primack04) imply instead
a soft spectrum ($\Gamma>2$), locating the IC peak below the observed
energy range ($<200\,\rm GeV$). In the following, we will discuss
both scenarios for the SED modelling, using the Phigh and Primack2004
curves as the two ends of the possible range of values for the
``Primack-type" shape (i.e., between the claimed EBL direct
measurements at few microns and the lower limits from galaxy counts).

\section{SED Modelling} \label{sect:dis}

\begin{figure*}
\includegraphics[angle=0,scale=1]{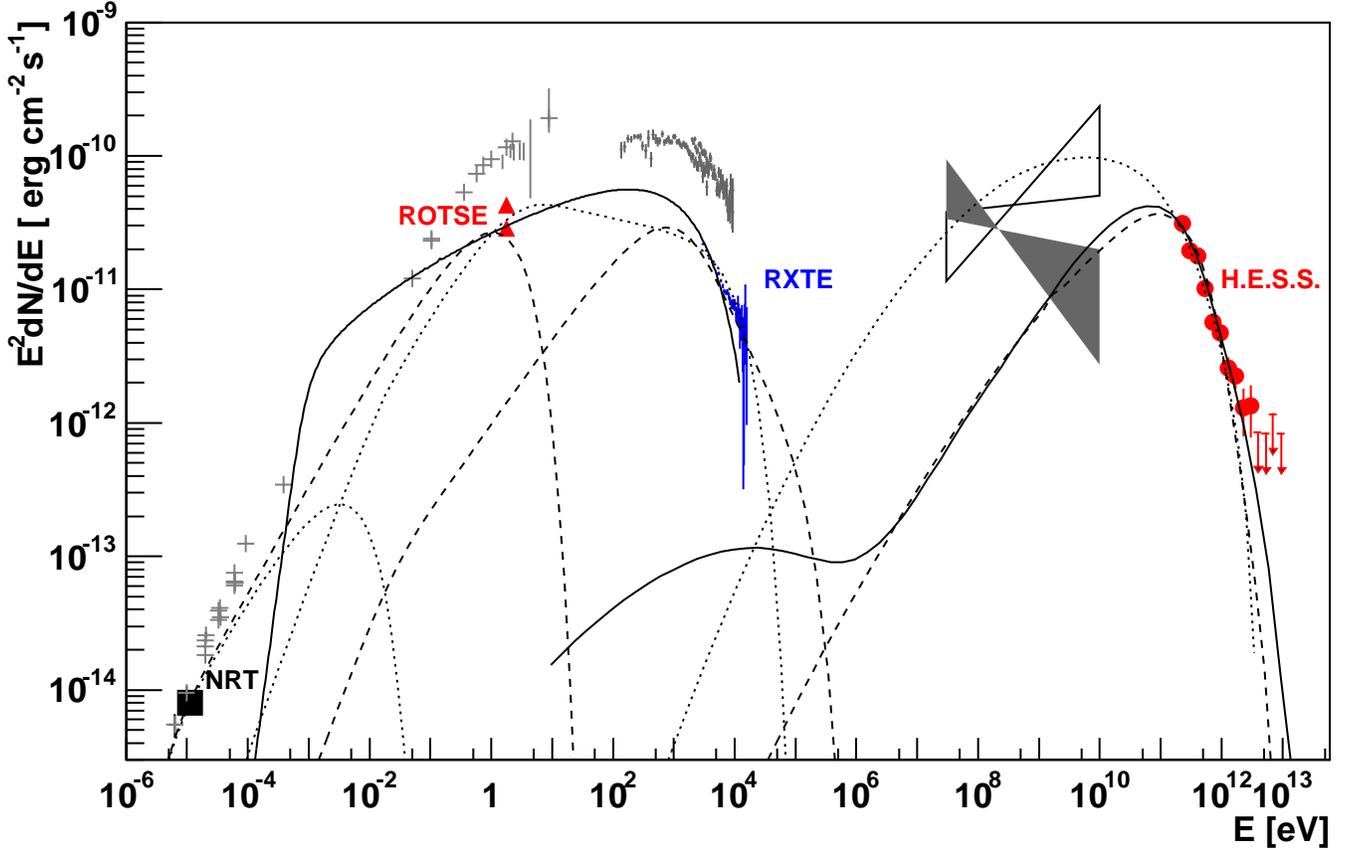}
\caption{Spectral energy distribution of PKS2155$-$304. Only
  simultaneous measurements are labeled. Non contemporaneous data are
  in grey symbols. The \hess\ spectrum is derived from October and
  November 2003 data (\textit{filled circles}) as is the \xte\
  spectrum. The NRT radio point (\textit{filled square}) is the
  average value for the observations carried out during this
  period. The two triangles are the highest and lowest ROTSE
  measurements for the Oct-Nov observations. Archival SAX data
  represent the high state observed in 1997 from \cite{chia}. Archival
  EGRET data are from the third EGRET catalogue (shaded bowtie),
  \cite{hart}) and from a very high $\gamma$-ray state described in
  \cite{vest2} (open bowtie). 1995 U B V R I data are from
  \cite{paltani} where the mean flux and flux deviation is used
  here. IUE data from \cite{urr} at 2800 \AA\ and 1400 \AA\ are
  included here with an error bar corresponding to the spread of the
  measured fluxes. Other data are NED archival data. The solid line is
  the hadronic model described in section \ref{hadrons}. The dotted
  and dashed lines are the same leptonic model with different
  assumptions described in section \ref{meudon}: the dotted line
  assumes a common origin for the optical and X-ray synchrotron
  emission, while the dashed line is the case where the optical
  emission emanates from the VLBI core; for both cases the lowest
  frequency hump is the predicted VLBI core emission. All the VHE
  emission in the models are absorbed according to the Primack04 model
  discussed in section \ref{ebl}.}
\label{fig:f7}
\end{figure*}

The broadband spectral morphology of \pks\ is typical of the BL Lac
type, with a double-humped structure in $\nu F_\nu$ representation,
exhibiting a low-energy and a high-energy component. Its broadband
emission is usually attributed to emission from a beamed relativistic
jet, oriented in a direction close to the line of sight
(\cite{begel,blandf}). The spectral energy distribution in units of
power per logarithmic bandwidth $E^2 {\rm d}N/{\rm d}E$ versus energy
$E$ is shown in Figure \ref{fig:f7}. The EGRET measurements, between
the \hess\ and \xte\ points, are from the third EGRET catalog
(\cite{hart}) and from a very high $\gamma$-ray state described in
\cite{vest2}. There is a difference in spectral states, since in the
former case the power law photon index is $1.71\pm0.24$ whereas it is
$2.34\pm0.26$ in the latter which most likely consists of a mix of low
and high activity state observations. The historical EGRET spectra are
therefore unlikely to represent the state of \pks\ during the campaign
presented here and are not used to put stringent constraints on the
modelling. Considering the archival data and the steep X-ray spectrum
in Figure \ref{fig:f7}, the peak of the low-energy component occurs in
the 2--$2000\,\rm eV$ range. The archival BeppoSAX data from a high
state analyzed by \cite{chia}, and represented here above our \xte\
data, show a peak at $\approx$ 0.1 keV. The absorbed VHE peak location
is clearly below $300\,\rm GeV$, with its exact location depending on
the spectrum in the EGRET range.

Whereas the current models seem
to agree that the low-energy component is dominated by synchrotron
radiation coming from nonthermal electrons emitted in collimated jets,
the high-energy emission is assumed to be either inverse Compton
scattering off the synchrotron photons (Synchrotron Self-Compton, SSC,
see e.g. \cite{maraschi,bicknell}) or by external photons (see
e.g. \cite{siko}). This kind of leptonic model will be discussed in
section \ref{meudon}. A hadronic origin of the VHE emission using the
Synchrotron-Proton Blazar (SPB) model with a dominating proton
synchrotron component at high energies in a proton-electron plasma is
also able to produce a double humped SED and is discussed in section
\ref{hadrons}. The lack of correlation between the RXTE and H.E.S.S.
fluxes (and possibly also between the optical and the VHE emission)
within the small variability range may indicate a different spatial
origin, or a different underlying particle distribution. In the
proton synchrotron model a lack of correlated variability between
$\gamma$-ray emission and the low energy electron synchrotron component
could arise if the electrons and protons are not co-accelerated.

The high-energy component above $\approx$ 100~GeV is attenuated
by interactions with the EBL and is a lower limit for the intrinsic
spectrum. The energy budget in X-rays
and VHE $\gamma$-rays is comparable, though the maximum output at the
peak energy in the high-energy component is likely to be lower than
that in the lower-energy component. Interpolating between the high-state EGRET
archival data and VHE data would lead to a maximum located above $10\,\rm
GeV$, which is surprising since the observations reported here
indicate a low state. Extrapolating the EGRET catalogue spectrum to
VHE energies with a power law falls below the \hess\ data and
therefore requires two inflexion points in the SED. Simultaneous
observations in the MeV-GeV range with the upcoming satellite GLAST
will be crucial to constrain the high-energy component shape.

\subsection{Doppler boosting and synchrotron/Compton derived parameters}

The electromagnetic emission in blazars is very likely to be
Doppler-boosted (or beamed) toward the observer. In the radio regime,
the evidence for Doppler boosting in \pks\ comes from superluminal
expansions observed with VLBI (\cite{piner}). Relativistic beaming is
also required in order to avoid absorption of GeV photons by X-ray
photons via the ${\rm e}^+/{\rm e}^-$ pair-production process (see e.g.
\cite{mara}). It is thus possible to use the $\gamma$-ray
variability to establish a limit for the Doppler factor $\delta$, with
$\delta$ defined in the standard way as
$[\Gamma(1-\beta\cos\theta)]^{-1}$, where $\Gamma$ is the bulk Lorentz
factor of the plasma in the jet, $\beta = v/c$, and $\theta$ is the
angle to the line of sight.

Following \cite{dondi}, the size of the $\gamma$-ray emission zone $R$
is derived from the time-scale of variability $t_{\rm var}$ (supposing
the timescale of intrinsic variability is negligible compared to the
light crossing time) by $R \leq ct_{\rm var}\delta/(1+z)$. In this
case (assuming a time scale $t_{\rm var} \approx 2$ ks from section
\ref{sect:res}) the size of the emission region is
\begin{equation}
R\delta^{-1}\leq 5\times 10^{13}\, {\rm cm}.
\label{eq:eq2}
\end{equation}

 Since for $\gamma$-rays in the $300\,\rm GeV$--$3\,\rm TeV$ range the
 opacity $\tau_{\gamma\gamma}$ cannot significantly exceed 1, independently of
 the emission mechanisms, another constraint on the minimum value of
 the Doppler factor can be derived by estimating how much Doppler
 boosting is necessary for photons with observed energy $E_\gamma$ to
 escape from a source with radius $R$ and a flux density $F(E_t)$,
 where $E_t=(m_ec^2)^2\delta^2/E_\gamma(1+z)^2 $. At this point one can follow
 \cite{mattox} and especially Eq. 3.7 in \cite{dondi}, writing that
 the opacity is
\begin{equation}
\tau_{\gamma\gamma} \approx \frac{\sigma_{\rm T}}{5} \frac{1}{hc}
\frac{d_L^2}{R} \frac{1}{\delta^3(1+z)} F(E_t).
\label{eq:eq5}
\end{equation}
Imposing that $\tau_{\gamma\gamma}<1$ for photons with observed energy
$E_\gamma=1\,\rm TeV$ yields a lower limit on $\delta$ for given $R$
that can be derived numerically from the observed SED. For \pks\ this
yields \begin{equation}
R^{-1}\delta^{-6.4} \leq 5.6\times10^{-24}\, {\rm cm}^{-1}.
\label{eq:eq6}
\end{equation}
The minimum allowable boost $\delta$ comes from combining
Eq. \ref{eq:eq2} and Eq. \ref{eq:eq6} and yields
\begin{equation}
\delta\geq19
\label{eq:eq7}
\end{equation}
which is higher than the limit obtained in a similar way in
\cite{tav98}, but at the lower end of the range usually obtained from
SSC modelling as in \cite{kata}. It is useful to stress that this
constraint is valid under the usual assumption that the region
emitting the SED optical flux (i.e., the target photons) is cospatial
with (or at least embeds) the high energy emitting region. The models
used later in this paper are therefore essentially single zone models.

If the observed X-rays are synchrotron radiation from nonthermal
electrons then the mean observed energy $<E>$ of an electron with
Lorentz factor $\gamma_{\rm e}$ is given by 
$$<E> \approx \frac{\delta}{1+z}\frac{21\hbar\gamma_{\rm e}^2qB}{15\sqrt{3}m_e}. $$ 
Using $<E>=10\,\rm keV$ here yields 
\begin{equation}
B\delta\gamma_{\rm e}^2 = 1.1\times10^{12}\, {\rm G}.
\label{eq:eq8}
\end{equation}
In the SSC scenario the same electrons can Comptonize ambient photons
up to the VHE regime, thus
\begin{equation}
\frac{\delta}{1+z}\gamma_{\rm e} m_ec^2 \geq 3\, {\rm TeV.}
\label{eq:eq9}
\end{equation}
Combining Eq. \ref{eq:eq8} and Eq. \ref{eq:eq9} yields
\begin{equation}
B\delta^{-1} \leq 1.1\times10^{12}\, \rm{G}\times \left(
\frac{3\,{\rm TeV}}{m_ec^2}\right)^{-2}(1+z)^{-2}.
\label{eq:eq10}
\end{equation}
A numerical application yields
\begin{equation}
B\delta^{-1} \leq 0.03\, {\rm G}
\label{eq:eq11}
\end{equation}
and hence the Lorentz factor is constrained to $\gamma_{\rm e}\geq
1.3 \times 10^6$ which is higher than that calculated as above for
Mkn~421 by \cite{taka} who derived $\gamma_{\rm e}>5\times10^5$ (and
$B\approx 0.2$~G) but lower than $\gamma_{\rm e}\approx 10^7$ found
in a similar way in 1ES1959$+$650 (\cite{giebels}).

The X-ray data presented above imply that the X-ray spectrum of \pks\
hardens as the source brightens. This is often measured in BL Lac
objects; a hardening of the spectrum when flares occur, and a blueward
shift of the peak of the synchrotron emission $\nu_{\mathrm{sync}}$
(and presumably higher energy inverse-Compton emission) by factors
that can be as large as 100 were measured in the cases of Mkn~501
(\cite{pian}), 1ES 1426$+$428 and PKS 0548$-$322 (\cite{costa}). In
the case of PKS 2005$-$489 (\cite{perl}), a more moderate shift of a
factor of 3 or less of the synchrotron emission was found. The
archival data suggest that $\nu_{\mathrm{sync}}$ lies in the UV band
for \pks, but no data were taken simultaneously in this campaign at
that wavelength.

The lack of indication for correlated X-ray/VHE variability does not
 imply that \pks\ behaves differently from VHE sources such as Mkn~421,
 for which VHE/X-ray correlation has been established on a much
 higher variability basis (see e.g. \cite{cui04}) with dynamical
 ranges of 30 in both energy bands. Limiting those observations to the
 same dynamical range as observed here would not allow any claim for
 correlation. Future observations of \pks\ with a higher variability
 amplitude would bring more insight into this.

\subsection{Leptonic interpretation}
\label{meudon}

Interpretation with a single zone SSC model of the SED of \pks\ has
already been proposed in the literature using two different
assumptions. In \cite{kata} the low energy tail of the SSC model is
used to account for the low-energy component up to the optical in the
SED. That component is decomposed into two sub-components by
\cite{chia} where the radio to optical emission has another origin
than the X-rays, which are assumed to come from the jet. These two
different interpretations are used here in the context of the leptonic
model described in \cite{katar1} which has already been applied to
Mkn~501 and Mkn~421 (\cite{katar2}). To constrain this model, only the
simultaneous data are used, since the archival data reported in the
SED of Figure \ref{fig:f7} are likely to not represent the state of
this source (note for example the difference in optical flux and the ROTSE
measurement). 

When using the Primack04 absorption, the model used here can reproduce
the X-ray through VHE part of the SED, but the \hess\ spectrum
constrains it such that the radio measurement can not be included in
the synchrotron bump predicted by the single-zone model. As for
Mkn~421 and Mkn~501, adding a more extended component than the VHE
emitting zone can provide an explanation for this. The origin is
probably the compact VLBI core which has a radio core to lobe ratio of
$\approx 1$ (\cite{laurent,piner}) and a typical size of $10^{18}\,\rm cm$,
more than two orders of magnitude larger than the VHE emitting
zone. This VLBI feature dominates the spectrum at low energy and is
included in the SED modelling here. An uncertainty remains which is the
high frequency cutoff of this VLBI component. The host galaxy
contribution to the optical flux is estimated to be $\approx
10^{-11}\,\rm erg\,cm^{-2}\,s^{-1}$, deduced from the magnitudes
given by KFS98 and assuming a low-redshift solar
metallicity elliptical galaxy of age equal to $13\,\rm Gyr$ ($R-H=2.4$),
corresponding to a mass of $5\times10^{11}M_{\odot}$ (\cite{fioc}). So
even at the measured low activity state of \pks\ the host galaxy is
not contributing much in the optical range.

\begin{table}[bt]
\begin{center}
\begin{tabular}{|c|c|c|c|} \hline
Parameter & Model 1 & Model 2 & Model 2 Phigh\\ \hline
$R_{\rm blob}$(10$^{15}$cm) & 6.5 & 1.5 & 1.5\\
$B$(G) & 0.15 & 0.25 & 0.02\\
$\delta_{\rm b}$ & 25 & 25 & 50\\
$K$(cm$^{-3}$) & $160$ & $2.0\times10^3$ & 240\\
$\gamma_{\rm break}(10^3)$ & 7.5 & 100 & 300\\
$\gamma_{\rm max}(10^5)$ & 3.8 & 9.0 & 5.0\\
$\alpha_1$ & 1.4 & 1.7 & 1.6\\
$\alpha_2$ & 3.2 & 4.65 & 4.5\\
$R_{\rm jet}(10^6{\rm cm})$ & 1.0 & 1.0 & 1.0\\
$L_{\rm jet}({\rm pc})$ & 55 & 55 & 55\\
$\delta_{\rm jet}$ & 2 & 2 & 2\\
$K_{\rm jet}$(cm$^{-3}$) & 40 & 40 & 40\\
$B_{\rm jet}$(G) & 0.04 & 0.04 & 0.04\\
$\gamma_{\rm break,jet}(10^3)$ & 2.5 & 45 & 45\\
\hline
\end{tabular}
\end{center}
\caption{Parameters for the SED fit with the assumption that the
  optical and X-ray emission are part of the jet synchrotron emission
  (Model 1 in the table, dotted line in Fig. \ref{fig:f7}) or that the
  optical emission emanates from the VLBI emission zone (Model 2 in the table,
  dashed lines in the same Figure). The parameters are described in
  \cite{katar1}.}
\label{tab:meudon}
\end{table}

The ROTSE measurement can be ascribed here to either the high-energy
tail of the VLBI component or to the synchrotron part of the SSC
model. Assuming a common origin for the X-rays and the optical emission,
and using a variability time scale of 0.1 d to constrain the emitting
zone, the model tends to predict a high IC flux, as shown in Figure \ref{fig:f7}. 
However, the lack of correlation between the X-rays and
the optical emission in our measurements - also suggested by
\cite{domi} based on less sensitive RXTE/ASM measurements - indicates that the optical emission could originate from the VLBI
component, which is modelled by a slight increase in the maximal
Lorentz factor of the emitting electrons. This in turn lessens the
constraint on the simultaneous SSC fit of the X-ray and VHE part and
allows a better fit of the VHE spectrum for smaller sizes of the
emitting zone. Detailed parameters of the two hypotheses are given in
Table \ref{tab:meudon}.

\begin{figure}
{\includegraphics[angle=0,scale=0.45]{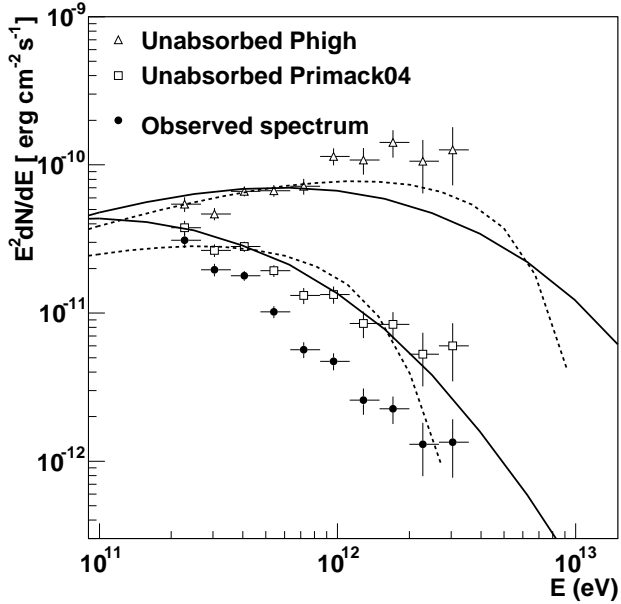}}
\caption{Estimations of the intrinsic \pks\ VHE spectrum, corrected
  for either the Phigh or Primack04 models, along with the associated
  intrinsic (i.e. before absorption) SSC model (dashed lines) and SPB
  model (solid lines).}
\label{fig:fin}
\end{figure}

If the absorption correction is well described by the Phigh model, the
slope of \hess\ data at high energy implies that the peak of the TeV
emission bump is located above $4\,\rm TeV$ (or $10^{27}\,\rm
Hz$). Such a high frequency peak emission imposes a strong constraint
for the single-zone SSC scenario, especially since the peak of the
synchrotron bump has to remain below $1\,\rm keV$ (or $10^{17.5}\,\rm
Hz$) as required by the slope of the RXTE data. High values of both
the jet Doppler factor and the maximal Lorentz factor of radiating
particles are required to reach the necessary energy for the IC bump,
that is $\gamma_{\rm max} \delta > 8 \times 10^6$. On the other hand,
to keep the synchrotron peak below $1\,\rm keV$ imposes an upper limit
to the magnetic field. Within these constrains, the best fit we obtain
is shown in Figure \ref{fig:fin}. We can note that none of the high
energy tails are well accounted for. The set of parameters for the
best fit is given in Table \ref{tab:meudon}. This fit marginally
reproduces the observed X-ray and $\gamma$-ray data, but is not as
satisfactory than that obtained with the Primack04 absorption
correction, and in any case it is impossible to take into account the
ROTSE optical point. The main changes in parameters between the two
fits consist in enhancing the boosting, which then becomes quite
extreme, while reducing the density and magnetic field for the Phigh
absorption correction.

The constraints on $\delta$ derived here from either simple opacity
  arguments or from the one-zone model parametrisation of the SED are
  in the range of Doppler factors usually derived with such
  assumptions or models for other VHE emitters. As pointed out by
  \cite{chiab} such high values are however at odds with attempts to
  unify the BL Lac population with the family of FR~I sources
  (\cite{urry2}), the latter being possibly an unbeamed since off-axis
  viewed case of the former. The same authors suggest that models
  where velocity structures in the jet, such as the ``spine-sheath''
  model (see e.g. \cite{sol}) or the decelerating flow model
  (\cite{georg,georg2}) allow lower bulk Lorentz factors. Another
  option from \cite{pellet} is to cope with the pair creation
  catastrophe implied by smaller Doppler factors in their ``two-flow''
  solution. Comparing the SED with such models, which make the BL Lac
  - FR-I connection more plausible, is an interesting task but beyond
  the scope of this paper.

\subsection{Hadronic models}
\label{hadrons}
Generally, the leptonic models constitute the preferred concept for
TeV blazars, essentially because of two attractive features: {\em (i)}
the capability of the (relatively) well understood shocks to
accelerate electrons to TeV energies (\cite{sikomad,pelletier}) and
{\em (ii)} the effective conversion of the kinetic energy of these
relativistic electrons into the X-ray and VHE $\gamma$-ray emission
components through the synchrotron and inverse Compton radiation
channels. The hadronic models are generally lack these
virtues. They assume that the observed $\gamma$-ray emission
is initiated by accelerated protons interacting with ambient matter
(\cite{bednar,laor,pohl}),
photon fields (PIC model, \cite{mannh}), magnetic fields
(\cite{ahaa}) or both (\cite{muck1}).

The models of TeV blazars involving interactions of protons with
photon and B-fields require particle acceleration to extreme energies
exceeding $10^{19}\,\rm eV$ which is possible if the acceleration
time is close to $t_{\rm acc}=r_{\rm g}/c$ ($r_{\rm g}$ is the
gyro-radius). This corresponds
(independent of a specific acceleration mechanism) to the maximum
(theoretically possible) acceleration rates (\cite{ahaa}) which
can only be achieved by the conventional diffusive shock acceleration in
the Bohm diffusion regime. 

On the other hand, the condition of high
efficiency of radiative cooling of accelerated particles requires
extreme parameters characterizing the sub-parsec jets and their
environments, in particular very high densities of the thermal plasma,
radiation and/or B-fields. In particular, the proton-synchrotron models 
of TeV blazars require highly magnetized ($B \gg 10\,\rm  G$) 
condensations of $\gamma$-ray emitting clouds containing Extremely
High Energy (EHE) protons, where the magnetic pressure dominates over
the pressure of relativistic protons (\cite{aha20}).

Below we use the hadronic SPB model
(\cite{muck1,muck2}) to model the average spectral energy distribution
(SED) of \pks\ in October-November 2003. A detailed description of
the model, and its implementation as a (time-independent)
Monte-Carlo/numerical code, has been given in \cite{mucke} and
\cite{reimer}.

Considering the rather quiet activity state of \pks\ in
 Oct-Nov. 2003, we use the 3EG catalog spectrum, since it is the best
 determined EGRET spectrum from this source to date, as an upper limit
 for modelling purposes.

 Flux variability provides an upper limit for the size of the
 emission region. To allow for a comparative study between leptonic
 and hadronic models we fix here the comoving emission region to $R
 \sim c t_{\rm var} \delta = 5\times 10^{13}\delta\,\rm cm$ deduced
 from the X-ray variability. We assume that the optical through
 X-ray emission and the $\gamma$-ray output stem from the same region of
 size $R$.

 A reasonable model representation for the simultaneous data assuming
 a Primack04 model for the VHE absorption is found for the following
 parameters: magnetic field $B=40\,\rm G$, Doppler factor $\delta=20$,
 injection electron spectral index $\alpha_{\rm e}=1.6$, assumed to be
 identical to the injection proton spectral index $\alpha_{\rm p}$,
 maximum proton energy of order $\gamma_{\rm p,max} \sim4\times
 10^{9}$, e/p-ratio of 0.15 and a near-equipartition proton energy
 density of $u_{\rm p}=27\,\rm erg\,cm^{-3}$. The required total jet
 power is of the order $L_{\rm jet}\sim 1.6\times10^{45}\,\rm
 erg\,s^{-1}$. When using the Phigh EBL model, a reasonable
 representation of the data may be achieved by increasing the maximum
 injected proton energy to $\gamma_{\rm p,max}=10^{10}$ and
 simultaneously increasing the e/p-ratio to 0.24, while all other
 parameters remain unchanged. Note that here the maximum proton
 gyro-radius approaches the size of the emission region.
 Alternatively, a doubling of the magnetic field to $80\,\rm G$
 together with an increase of $\gamma_{\rm p,max}$ to $~8\times 10^9$
 and a e/p-ratio of unity (leading to $u_B\approx 50 u_{\rm p}$)
 represents the SED-data equally well. In conclusion, none of the
 "Primack-type" EBL models can explicitly be ruled out in the
 framework of the SPB-model by the \hess\ data presented here. In all
 cases, proton synchrotron emission dominates the (sub-)TeV radiative
 output. Depending on the Doppler factor, part of the proton
 synchrotron radiation produced may be reprocessed to lower
 energies. Contributions from the muon and pion cascades are always
 lower than the proton synchrotron component. The low energy
 component is dominated by synchrotron radiation from the primary
 electrons, with a negligible contribution of synchrotron radiation
 from secondary electrons (produced by the $\rm p$- and
 $\mu^\pm$-synchrotron cascade).

On the other hand, the synchrotron radiation of secondary electrons
 resulting from interactions of VHE $\gamma$-rays with external
 low-energy photons with a modest $\gamma \gamma \rightarrow {\rm e}^+/{\rm e}^-$
 opacity ($\tau_{\gamma \gamma} \leq 1$) may lead to significant
 X-ray emission with a luminosity comparable to the luminosity of the
 primary VHE emission (\cite{aha20}).

 Models involving meson production inevitably predict neutrino emission
 due to the decay of charged mesons. The SPB-model for \pks\
 explains the high energy emission dominantly as proton synchrotron
 radiation, making the neutrino flux completely negligible.

\section{Conclusions}

This paper reports multi-wavelength observations of the BL Lac object
\pks\ in 2003. Although the source appeared variable in the VHE,
X-rays and optical bands, the latter two indicate that \pks\ was in a
state close to historically low measurements, even though it
was easily detectable by \hess\ in all nights of observations since
the beginning of the detector operation (see AH04 for the observation
history up to August 2003). An extreme case of
VHE variability shows a peak-to-peak increase of a factor of $2.5\pm0.9$
in 0.09 d. Variability on the timescale of a few ks in the 2--$10\,\rm
keV$ band and of the order of $0.1\,\rm d$ for energies $>300\,\rm
GeV$ were observed by \xte\ and \hess\ ~The X-ray data show a
correlation of the flux with the spectrum, which becomes harder when
the source is brighter. At the level of the simultaneously observed
 modest variability, no correlation between the VHE $\gamma$-rays, the X-rays
and optical was seen. Observations with greater variability and better
coverage are needed before it can be asserted that the VHE/X-ray
pattern in \pks\ is different from other known VHE-emitting AGN. Since
the source was in a low emission state in both the optical and X-rays
compared to archival measurements, this lack of correlation has yet to
be established for a higher emission state. Simultaneous observations
in the X-rays/optical band and VHE $\gamma$-rays had never previously
been performed on this scale. Its continual VHE detection makes \pks\
unique in the TeV BL Lac category, and probably indicates that \hess\
has achieved the sensitivity level where it can detect the quiescent
state of \pks\ at any time. A time-averaged energy spectrum is
determined for the 2 observation periods and fits to a power law
($\Gamma = 3.37\pm0.07$) in the VHE $\gamma$-rays, and to a broken
power law ($\Gamma_L=2.81\pm0.05$, $E_{b}=4.9\pm0.8\,\rm keV$,
$\Gamma_H=2.95\pm0.04$) in the X-rays.

A comparison of the intrinsic spectrum with predictions from existing
one-zone leptonic and one-zone hadronic models is attempted to give a
plausible estimation of underlying physical parameters. The values of
the parameters are in line with those inferred for other VHE-emitting
blazars. In these models the VHE emission is attenuated according to
two different EBL levels. This changes mainly the Doppler boosting in
the leptonic model, but the high level EBL decreases the agreement
significantly. In the hadronic model, the maximum injected proton
energy can be changed to accomodate different EBL levels and can
therefore not significantly constrain any of the EBL models used
here.

\begin{acknowledgements}
  The authors would like to thank the anonymous referee for his
  correction and useful comments that improved this paper. The support
  of the Namibian authorities and of the University of Namibia in
  facilitating the construction and operation of H.E.S.S.  is
  gratefully acknowledged, as is the support by the German Ministry
  for Education and Research (BMBF), the Max-Planck-Society, the
  French Ministry for Research, the CNRS-IN2P3 and the Astroparticle
  Interdisciplinary Programme of the CNRS, the U.K. Particle Physics
  and Astronomy Research Council (PPARC), the IPNP of the Charles
  University, the South African Department of Science and Technology
  and National Research Foundation, and by the University of Namibia.
  We appreciate the excellent work of the technical support staff in
  Berlin, Durham, Hamburg, Heidelberg, Palaiseau, Paris, Saclay, and
  in Namibia in the construction and operation of the equipment.

The authors acknowledge the support of the ROTSE III collaboration and
the sharing of observation time with the Australian ROTSE IIIa
telescope operated by A. Phillips and M.C.B. Ashley from the School of
Physics, Department of Astrophysics and Optics, University of New
South Wales, Sydney, Australia. Special thanks also to R. Quimby from
the University of Texas for providing tools for data-reduction. 

H. Sol and C. Boisson thank K. Katarzy\'nski for his SSC code.

This research has made use of the NASA/IPAC Extragalactic Database
(NED) which is operated by the Jet Propulsion Laboratory, California
Institute of Technology, under contract with the National Aeronautics
and Space Administration. We thank the \xte\ team for their prompt
response to our ToO request and the professional interactions that
followed.
\end{acknowledgements}

\end{document}